%Paper: 9205004
%From: Dennis Clougherty <dpc@sbphy.physics.ucsb.edu>
%Date: Tue, 5 May 92 13:21:51 -0700

%%%%%%%%%%%%%%%%%%%%%%%%%%%%%%%%%%%%%%%%%%%%%%%%
%                                              %
% The TeX file with text and figure captions   %
% begins here.  A postscript file for          %
% figure 1 follows. The macro reforder is      %
% used. The figure is                          %
% separated by a heading like this one.        %
% It should be separated into its own          %
% file and then printed separately.            %
%                                              %
%%%%%%%%%%%%%%%%%%%%%%%%%%%%%%%%%%%%%%%%%%%%%%%%

\magnification=\magstep1

\hsize=6.5truein
\vsize=8.9 truein
\baselineskip=\normalbaselineskip \multiply\baselineskip by
2
\def\beginlinemode{\endmode\begingroup\parskip=0pt
\obeylines\def\\{\par}\def\endmode{\par\endgroup}}

\def\beginparmode{\endmode\begingroup
\def\endmode{\par\endgroup}}

\def\endpage{\vfill\eject}

\def\raggedcenter{\leftskip=4em plus 12 em
\rightskip=\leftskip \parindent=0pt \parfillskip=0pt
\spaceskip=.3333em \xspaceskip=.5em \pretolerance =9999
\tolerance=9999 \hyphenpenalty=9999 \exhyphenpenalty=9999 }

\let\endmode=\par{\obeylines\gdef\
{}}

\overfullrule=0pt

\medskipamount=7.2pt plus2.4pt minus2.4pt

\parskip=\medskipamount

\def\refto#1{$^{#1}$}

\gdef\refis#1{\indent\hbox to 0pt{\hss#1.~}}

\def\ref#1{Ref. #1}			% 	for inline references
\def\Ref#1{Ref. #1}			% 	ditto
\def\eq#1{Eq. (#1)}			%	ditto
\def\Eqs#1{Eqs. (#1)}			%	ditto
\def\eqs#1{Eqs. (#1)}			%	ditto
\def\ie{{\it i.e.,\ }}

\def\oneandahalfspace{\baselineskip=\normalbaselineskip
  \multiply\baselineskip by 3 \divide\baselineskip by 2}
\newcount\firstpageno
\firstpageno=2
\footline={\ifnum\pageno<\firstpageno{\hfil}\else{\hfil\folio\hfil}\fi}

\def\head#1{			% Head;  NOTE enclose the text in {}
  \filbreak\vskip 0.5truein	%  e.g., \head{I. Introduction}
  {\immediate\write16{#1}
   \raggedcenter \uppercase{#1}\par}
   \nobreak\vskip 0.25truein\nobreak}

\def\references			% Begin references -- basic format is Phys Rev
  {\head{References}		% I.e., volume, page, year (space after commas).
   \beginparmode
   \frenchspacing \parindent=0pt \leftskip=1truecm
   \parskip=8pt plus 3pt \everypar{\hangindent=\parindent}}

\def\body			% Begin text body;  can be used to end
  {\beginparmode}		% \title, \author, \affil, \abstract,
				% \reference, or \figurecaption modes

\def\endreferences{\body}

\def\etal{{\it et al.}}
%reforder.tex
\catcode`@=11
\newcount\r@fcount \r@fcount=0
\newcount\r@fcurr
\immediate\newwrite\reffile
\newif\ifr@ffile\r@ffilefalse
\def\w@rnwrite#1{\ifr@ffile\immediate\write\reffile{#1}\fi\message{#1}}

\def\writer@f#1>>{}
\def\referencefile{%			  Stuff to write .REF file
  \r@ffiletrue\immediate\openout\reffile=\jobname.ref%
  \def\writer@f##1>>{\ifr@ffile\immediate\write\reffile%
    {\noexpand\refis{##1} = \csname r@fnum##1\endcsname = %
     \expandafter\expandafter\expandafter\strip@t\expandafter%
     \meaning\csname r@ftext\csname r@fnum##1\endcsname\endcsname}\fi}%
  \def\strip@t##1>>{}}

\def\citeall#1{\xdef#1##1{#1{\noexpand\cite{##1}}}}
\def\cite#1{\each@rg\citer@nge{#1}}	% Variable No. of args, separated by ","

\def\each@rg#1#2{{\let\thecsname=#1\expandafter\first@rg#2,\end,}}
\def\first@rg#1,{\thecsname{#1}\apply@rg}	% each@ag is a general purpose
\def\apply@rg#1,{\ifx\end#1\let\next=\relax%	  variable no. of arg. macro.
\else,\thecsname{#1}\let\next=\apply@rg\fi\next}% args separated by commas

\def\citer@nge#1{\citedor@nge#1-\end-}	% Check for M-N range (M and N numbers)
\def\citer@ngeat#1\end-{#1}
\def\citedor@nge#1-#2-{\ifx\end#2\r@featspace#1 % Single argument
  \else\citel@@p{#1}{#2}\citer@ngeat\fi}	% M-N range of arguments
\def\citel@@p#1#2{\ifnum#1>#2{\errmessage{Reference range #1-#2\space is bad.}%
    \errhelp{If you cite a series of references by the notation M-N, then M and
    N must be integers, and N must be greater than or equal to M.}}\else%
 {\count0=#1\count1=#2\advance\count1
by1\relax\expandafter\r@fcite\the\count0,%
  \loop\advance\count0 by1\relax%	  Loop from M to N
    \ifnum\count0<\count1,\expandafter\r@fcite\the\count0,%
  \repeat}\fi}

\def\r@featspace#1#2 {\r@fcite#1#2,}	% Eat spaces at beginning or end of arg
\def\r@fcite#1,{\ifuncit@d{#1}%		  Cite individual reference
    \newr@f{#1}%
    \expandafter\gdef\csname r@ftext\number\r@fcount\endcsname%
                     {\message{Reference #1 to be supplied.}%
                      \writer@f#1>>#1 to be supplied.\par}%
 \fi%
 \csname r@fnum#1\endcsname}
\def\ifuncit@d#1{\expandafter\ifx\csname r@fnum#1\endcsname\relax}%
\def\newr@f#1{\global\advance\r@fcount by1%
    \expandafter\xdef\csname r@fnum#1\endcsname{\number\r@fcount}}

\let\r@fis=\refis			% Save old \refis, redefine
\def\refis#1#2#3\par{\ifuncit@d{#1}%      Use two params #2 #3 to strip blank
   \newr@f{#1}%
   \w@rnwrite{Reference #1=\number\r@fcount\space is not cited up to now.}\fi%
  \expandafter\gdef\csname r@ftext\csname r@fnum#1\endcsname\endcsname%
  {\writer@f#1>>#2#3\par}}

\def\ignoreuncited{%   redefine \refis if ignoring uncited references
   \def\refis##1##2##3\par{\ifuncit@d{##1}%
     \else\expandafter\gdef\csname r@ftext\csname
r@fnum##1\endcsname\endcsname%
     {\writer@f##1>>##2##3\par}\fi}}

\def\r@ferr{\endreferences\errmessage{I was expecting to see
\noexpand\endreferences before now;  I have inserted it here.}}
\let\r@ferences=\references
\def\references{\r@ferences\def\endmode{\r@ferr\par\endgroup}}

\let\endr@ferences=\endreferences
\def\endreferences{\r@fcurr=0%		  Save old \endreferences, redefine
  {\loop\ifnum\r@fcurr<\r@fcount%	  Loop over refnum and produce text
    \advance\r@fcurr by 1\relax\expandafter\r@fis\expandafter{\number\r@fcurr}%
    \csname r@ftext\number\r@fcurr\endcsname%
  \repeat}\gdef\r@ferr{}\endr@ferences}

% Save old \endpaper, redefine it to write parting message.

\let\r@fend=\endpaper\gdef\endpaper{\ifr@ffile
\immediate\write16{Cross References written on []\jobname.REF.}\fi\r@fend}

\catcode`@=12

\def\reftorange#1#2#3{$^{\cite{#1}-\setbox0=\hbox{\cite{#2}}\cite{#3}}$}

\citeall\refto		% These macros will generate citations
\citeall\ref		%
\citeall\Ref		%
%*****************************************************

\rightline{UCSBTH-92-06}
\rightline{cond-mat/9205004}

\null\vskip 3pt plus 0.2fill \beginlinemode
\raggedcenter {\bf Low Energy Behavior of Quantum Adsorption}

\vskip 3pt plus 0.2fill D.P. Clougherty and W. Kohn

\vskip 3pt plus 0.2fill {\sl Department of Physics\\
 University of California\\Santa Barbara, CA  93106--9530}

\vskip 3pt plus 0.3fill \beginparmode
\oneandahalfspace
\narrower We present an exact solution of a 1D model: a particle of
incident energy $E$ colliding with a target which is a 1D harmonic
``solid slab'' with $N$  atoms in its ground state; the Hilbert space of
the target is restricted to the ($N+1$) states with zero or one phonon
present. For the case of a short range interaction, $V(z)$, between the
particle and the surface atom supporting a bound state, an explicit
non-perturbative solution of the collision problem is presented. For
finite and large $N$, there is no true sticking but only so-called
Feshbach resonances. A finite sticking coefficient  ${\sl s}(E)$ is
obtained by introducing a small phonon decay rate $\eta$ and letting
$N\to\infty$. Our main interest is in the behavior of ${\sl s}(E)$ as
$E\to 0$. For a short range $V(z)$, we find ${\sl s}(E)\sim E^{1/2}$,
regardless of the strength of the particle-phonon coupling. However, if
$V(z)$ has a Coulomb $z^{-1}$ tail, we find  ${\sl s}(E)\to\alpha$,
where $0 < \alpha < 1$. [A fully classical calculation gives ${\sl
s}(E)\to 1$ in both cases.] We conclude that the same threshold laws
apply to 3D systems of neutral and charged particles respectively.
\smallskip \noindent{PACS numbers: 68.10.Jy, 03.65.-w, 03.80.+r}

\endpage\beginparmode

The low energy behavior of the sticking coefficient ${\sl s}(E)$ of a
particle striking a surface and being trapped in a surface bound state
is still a matter of experimental\reftorange{nayak} {schlichting, doyle}
{mills}  and theoretical\reftorange{lj}{brenig, martin1, martin2}
{carraro} controversy. We consider here only a zero temperature target.
For a particle treated classically which interacts with a classical
elastic solid, it is known\refto{iche} that the sticking probability,
${\sl s}(E)$, tends to 1, as the incident energy of the particle $E\to
0$. In contrast, a particle whose motion is treated quantum mechanically
can have a dramatically different threshold behavior for sticking; it is
known\reftorange{lj}{brenig}{martin1} that for sufficiently weak
particle-phonon coupling, if perturbation theory is valid, a quantum
particle experiencing a short range interaction will have a sticking
probability vanishing as ${\sl s}(E)\sim E^{1/2}$ due to ``quantum
reflection.''

A number of experiments have attempted to explore the low energy region
where quantum effects should dictate the form of the threshold behavior
of ${\sl s}(E)$. Nayak \etal\refto{nayak}  found for $\rm ^4He$ atoms
striking a liquid-$\rm ^4He$ surface,  ${\sl s}\to 0$, in agreement with
the quantum prediction.  Similarly, results\refto{schlichting} of low
energy scattering of Ne from Ru(001) indicate the vanishing of ${\sl
s}(E)$ as $E\to 0$.  However, more recently, ${\sl s}(E)$ was
determined experimentally\refto{doyle} for ultra-low energy atomic H
striking a liquid-$\rm ^4He$ surface, and  did not appear to have
quantum threshold behavior. Further threshold behavior apparently
contrary to quantum predictions was found for positronium (Ps)  on $\rm
Al(111)$ surfaces.\refto{mills} We shall return to these experiments at
the end.

Theoretically the threshold behavior of ${\sl s}(E)$ for a particle
coupled to a solid by a short range interaction was formally studied by
Brenig\refto{brenig} who expressed many-body effects in a non-local,
complex, energy dependent potential $U_{\rm eff}(r,r';E)$.  Assuming
that $U_{\rm eff}$  is also short range in $r$ and $r'$ and has a well
defined finite limit as $E\to 0$, he found ${\sl s}(0)=0$, in keeping
with quantum perturbation theory. The same conclusion\refto{boheim} was
reached for the important case of van der Waals interactions ($\sim
z^{-3}$ for large $z$).

Polarization effects due to virtual (non-resonant) phonon excitations
that are quantitatively important are neglected in Refs. \cite{brenig}
and  \cite{boheim}. Knowles and Suhl\refto{suhl} have shown that surface
polarization effects {\it increase} ${\sl s}(E)$ at low energies. As a
result of the polarization, the penetration of the particle's effective
wavefunction into the surface  region is increased. This  effect is
known to be in competition with quantum  reflection in the determination
of ${\sl s}(E)$ as $E\to 0$.

For the case of a charged particle which experiences a long range image
potential, Martin \etal\refto{martin1} calculated ${\sl s}(E)$ using
perturbation theory and concluded (mistakenly) that ${\sl s}(E)\propto
E^{1/4}$ for small $E$, so that ${\sl s}(0)=0$. Their subsequent
numerical  calculations using the time dependent Hartree approximation
indicated to them that ${\sl s}(0)  \ne 0$ if the particle-phonon
coupling $\lambda$ exceeded a critical value $\lambda_c$. They concluded
that when
 $\lambda > \lambda_c$, polarization effects
 dominate the quantum reflection.

We shall present results of an exactly solvable  one dimensional (1D)
model for quantum sticking. The model is sketched in Fig. 1a. It has an
external particle interacting with
 a 1D
 ``solid slab.'' All motions are constrained to 1D.  The surface atom
and the particle interact by a short  range potential  whose generic
form is sketched in Fig. 1b.  We take for the Hamiltonian of the
system\refto{lj&s} $$ {\cal H}={\cal H}_{ph}+{\cal H}_{p}+{\cal H}_{I},
\eqno(1) $$ where $$ \eqalign{ {\cal H}_{ph}&=\sum_{q}{\hbar \Omega_q
{a^\dagger}_q a_q},\cr {\cal H}_{p}&={P^2\over 2m}+V(z),\cr {\cal
H}_{I}&={1\over\sqrt{N}}\ \sum_q {{\sqrt{\hbar\over{M\Omega_q}}} \ \cos
\left({q a\over 2}\right) \ ({a^\dagger}_q + a_q)\ V'(z)};} \eqno(2) $$
${\cal H}_p$ is the Hamiltonian for the particle  moving in the static
potential, $V(z)$. ${\cal H}_{ph}$ is the Hamiltonian for the phonons in
the solid;   ${\cal H}_{I}$ contains the particle--phonon coupling; $m$
is the particle mass; $\Omega_q$ is the frequency of the phonon with
wavenumber $q$; and ${a^\dagger}_q$ and ${a_q}$ are phonon creation and
annihilation operators respectively. $N$ is the number of atoms in the
solid, $M$ is the mass of  a lattice atom, and $a$ is the equilibrium
lattice spacing.

The phonon Hamiltonian  describes a chain of $N$ atoms coupled
harmonically to their nearest neighbors and to  fixed sites as
illustrated  in Fig. 1. (The coupling to fixed sites, which introduces a
lower cut-off frequency $\Omega_c$, is  needed in this 1D model to
prevent a well-known infrared divergence which does not exist in higher
dimensions.) The interaction potential, $V$, is sufficiently deep so as
to support a bound state with energy $-E_b$.

We next restrict the Hilbert space to states with $0$ or $1$ phonon
present\refto{stiles2}. However we allow a  particle-lattice interaction
of arbitrary strength.

We expand the wavefunction using $0-$ and $1-$phonon eigenstates,
$\psi_0$ and $\{\psi_i\}$: $$ \Psi(z_1, z_2, \cdots, z_N, z)=
{\phi_0(z)\ \psi_0(z_1, z_2, \cdots, z_N)}+ \sum_{i=1}^{N} {\phi_i(z)\
\psi_i(z_1, z_2, \cdots, z_N)}, \eqno(3) $$ where $(z_1, z_2, \cdots,
z_N)$ are the positions of the chain atoms, $i$ labels its modes, and
$z$ is the position of the particle. The following coupled system of
equations results: $$ \eqalignno{ ({\cal H}_{p} -E)\
\phi_0(z)+\sum_{i=1}^N V_{0i}(z)\ \phi_i(z)&=0, &(4 a)\cr ({\cal
H}_{p}+\hbar\Omega_i -E)\ \phi_i(z)+V_{i0}(z)\ \phi_0(z)&=0, \ \ \
i=1,\cdots, N &(4 b)\cr } $$ where $$ V_{i0}(z)=V_{0i}(z)= {1\over\sqrt
N}\ \ {\sqrt{\hbar\over{M\Omega_i}}}\ V'(z)\ \cos ({q_i a  \over 2}),
\eqno(5) $$ $\phi_i$ is the wavefunction for the particle in the
i$^{th}$ channel. $\hbar\Omega_i$ is the excitation energy, $q_i$ is the
wavenumber of the $i^{th}$ mode, and $E$ is the total energy of the
system.

The coupled system of \eq{4} may be reduced to an effective one-particle
equation by first solving \eqs{4 b} for the $\phi_i$ in terms of
$\phi_0$. Substitution into \eq{4 a} gives a single equation for the
particle wavefunction in the elastic channel, $\phi_0$.

$$ \bigg({d^2 \over dz^2}+k^2-U(z)\bigg)\ \phi_0(z)-\int dz' \ U_{\rm
eff}(z,z';k)\ \phi_0(z')=0, \eqno(6) $$  where  $$ U_{\rm eff}(z,z';k)
\equiv\sum_{i=1}^N\ U_{0i}(z)\ U_{0i}(z')\ G_i(z,z') \eqno(7) $$ and  $$
G_i(z,z')={\Phi_b(z) \ \Phi_b(z')
\over{\omega_i-k^2-\epsilon_b}}+\int_0^\infty dk'\ {\Phi(z,k') \
\Phi(z',k') \over{\omega_i-k^2+k'^2-i\eta}}, \eqno(8) $$
$k\equiv\sqrt{2mE}/\hbar$,   $\omega_i\equiv{2m\over\hbar^2}\
\hbar\Omega_i$, and  $\epsilon_b\equiv{2m E_b \over\hbar^2}$.
$\Phi(z,k)$ and $\Phi_b(z)$ are normalized continuum and bound state
eigenfunctions respectively for the  static potential $U(z)$. $\eta\to
0^+$, so as to satisfy the outgoing boundary conditions for the
scattered particle.

The boundary/asymptotic conditions on $\phi_0$ are $$ \eqalignno{
\phi_0(&-\infty)=0 &(9 a)\cr
\phi_0(&z){\smash{\mathop{=}\limits_{z\to\infty}}} e^{-ikz}-R(k)
e^{ikz},&(9 b)\cr} $$ where $R(k)$ is the elastic reflection coefficient.
\Eqs{5} and (8) can be analytically solved for ${\sl s}(E)$.  Complete
details are given elsewhere.\refto{ck} We find that for {\it finite}
$N$, ${\sl s}(E)\equiv 0$ for all $E$. Only in the limit of $N\to\infty$
does sticking occur, and ${\sl s}(E)$ is given as  $$  {\sl
s}(E)={1\over k}\ {|P(k)|^2 |{\rm Im}\ {\cal I}(k)| \over |1-W(k){\cal
I}(k)|^2}  \eqno(10) $$ where  $$ W(k)=2\ ({m\over M})\  \int dz' dz'' \
\Phi_b(z')\ U'(z')\ {\cal G}(z',z'';k) \ U'(z'') \ \Phi_b(z''), \eqno(11)
$$ $$ {\cal I}(k)= \int_{\omega_c}^{\omega_m}d\omega \ {\rho(\omega)
\cos^2 \left({q(\omega) a\over 2}\right) \over{(k^2
+\epsilon_b-\omega+i\eta)\ {\omega}}} \eqno(12) $$ and $$
P(k)=-2ie^{i\delta(k)}\ \sqrt{2m\over M} \ \int dz' \Phi_b(z')\ U'(z')\
\chi_0(z';k); \eqno(13) $$ $\rho(\omega)$ is the density of vibrational
states per atom and ${\cal G}$ is a Green's function for the particle in
the potential, $U+U_{\rm pol}$, consisting of the static part and the
part due to virtual excitations of the particle-phonon system, exclusive
of virtual particle sticking; $\chi_0$ is  a solution in $U+U_{\rm pol}$,
subject to the boundary condition that it vanish as $z\to -\infty$ and
the asymptotic condition that it approach $\sin(kz+\delta)$ as
$z\to\infty$; $\delta$ is the phase shift resulting from $U+U_{\rm pol}$.

In the low energy regime, for a finite range $U$,
$\chi_0(z,k){\smash{\mathop{\longrightarrow}\limits_{k\to 0}}}\ k\
f(z)$. The factor $k$ is the manifestation of quantum reflection. The
result is that regardless of the coupling strength,  $$ {\sl
s}(E){\smash{\mathop{\sim}\limits_{E\to 0}}}\ E^{1/2} \eqno(14) $$ For a
neutral particle which asymptotically experiences a $z^{-3}$ potential,
the above result is still valid\refto{boheim}.

A charged particle asymptotically experiences a $z^{-1}$ potential. Here
(unlike for $z^{-3}$) the WKB approximation for  $\chi_0$ is valid for
all $k$ beyond a fixed $z$, and the particle experiences no quantum
reflection. In the low energy regime,
$\chi_0(z,k){\smash{\mathop{\longrightarrow}\limits_{k\to 0}}}\
\sqrt{k}\ g(z)$. We define as ${\sl s}_n(E)$ the sticking probability
for the bound state $n$. In lowest order perturbation theory in the
particle-phonon coupling, we find $$ {\sl
s}_n(E){\smash{\mathop{\longrightarrow}\limits_{E\to 0}}}\ \alpha_n
\eqno(15) $$ where $0 < \alpha_n <1$.

We must concern ourselves with the convergence of the infinite summation
resulting from sticking contributions from the infinite number of
Coulomb bound states. The amplitude of high lying bound states near the
surface behaves as $n^{-{3/2}}$ as for pure Coulomb wave
functions\refto{martin1}. Thus the square of the matrix elements
decrease as $n^{-3}$, insuring convergence of the summation. In fact,
most sticking occurs in the lowest bound state.

In relating quantum sticking to classical sticking we want to point out
{\underbar{two}} quite distinct quantum effects.

\item{1.} A Debye-Waller like effect: In quantum mechanics there is a
{\underbar{finite}} probability (even as ${N}{\rightarrow}{\infty})$
that {\underbar{no}} lattice vibrations are excited and hence the
particle is reflected. Thus under all circumstances ${s}({E})<{1}$. By
contrast, classically, in the case of an attractive particle-target
interaction, a {\underbar{finite}} amount of impact energy is delivered
to the target, even when ${E}{\rightarrow} {0}$, because of  the
particle's acceleration by the interaction potential. When ${N}
{\rightarrow}{\infty}$, some of this energy disappears to $z=-\infty$.
Thus for ${E}$ sufficiently small, ${E}<{E_{\rm min}}$, the particle
cannot escape and ${s}({E}) = {1}$.

\item{2.} Quantum reflection:  We consider first the particle striking a
rigid target in the classical  regime. The particle coming in with a
{\underbar{low}} velocity, $-{v_\infty}$, spends a time of the order
${t_{\rm res}}{\sim}{{2z_0}\over{\bar v}}$ in the interaction region,
where ${z_0}$ is the range of interaction and ${\bar v}$ is a mean speed
in the interaction region. As ${v_{\infty}}{\rightarrow} {0}$, $\bar v$
approaches a finite limit and the ratio  of the time spent by the
particle in the interaction region to the time  spent in a spatial
interval ${z_0}$ in the asymptotic region is $$ {{\rm P_i}\over{\rm
P_\infty}}~~{\equiv}~~{{\rm t_{res}}\over({z_0\over v_\infty})}~~
{\sim}~~\bigg({E\over\bar E}\bigg)^{1/2}, \eqno(16) $$ where ${\bar
E}\,{\sim}\,{1\over{2}}\,m {\bar v}^2$ is a typical kinetic energy in
the interaction region, when ${v_{\infty}}{\rightarrow} {0}$.

Now consider the problem quantum mechanically for small incident
energy. In the {\underbar{rigid}} target potential, assumed
sufficiently short range, the particle is described by a standing
wave, ${\chi_0}({z},{k})$, with the properties $$ \eqalignno{
{\chi_0}({z},{k}) &{\smash{\mathop{=}\limits_{z\to\infty}}} {\cal N}\
\sqrt{{2}\over{\pi}}\,\, {\sin}({kz}+{\delta^{\prime}}) &(17 a)\cr
{\chi_0}({z},{k}) &{\smash{\mathop{\sim}\limits_{z\to 0}}} {\cal N} \
({k}{z_0})\ {f}({z}) &(17 b)\cr } $$ where $\cal N$ is an (irrelevant)
normalization factor, ${\delta^{\prime}}$ is a phase shift, and
${f}({z})$ becomes independent of ${k}$ for small $k$ and is of order
1. (This is well-known from so-called effective range
theory\refto{goldberger} and can easily be checked for a square well
interaction potential backed by an infinite wall.)
 Thus the ratio of the probability of finding the particle in the
interaction region to the probability of finding it in an asymptotic
interval of length ${z_0}$ is $$ {{\rm P_i}\over{\rm P_\infty}}
\approx  {{{\int_0^{z_0}}[({kz_0}) {f}({z})]^2 {dz}}\over {z_0}}\,\,
{\approx}\,\,{k^2}{z_0^2}\,\,{\approx}\,\,{{2mz_0^2}\over{\hbar^2}}\,{E}
\eqno(18) $$ Note the power of ${E^1}$ compared to the classical
result, ${E^{1/2}}$: as ${E}{\rightarrow}{0}$ the quantum particle
spends less time in the interaction region than the classical
particle, by a power ${E^{1/2}}$. This is the so-called quantum
reflection. We can also note that for unit incident current, the
probability of a quantum particle being in the interaction region is
${\sim}{E^{1/2}}$. This is the physical origin of the sticking
threshold behavior of \eq{14}. For a charged particle, on the other
hand, one finds ${{\rm P_i}/{\rm P_\infty}}$ has the form of \eq{16},
as in the classical case; \ie there is no quantum reflection.

Very recent experiments with low energy neutral particles striking low
temperature targets have failed to find the threshold behavior of
\eq{14} which we expect because of quantum reflection. The sticking of
neutral H on liquid $^4$He film\refto{doyle} was analyzed by Carraro and
Cole\refto{carraro} using Ref. \cite{boheim}. They conclude that the
energy ($\sim 10^{-8}$ eV) was \underbar{still} too high to observe
quantum reflection in this system.

An ingenious experiment by Mills \etal\refto{mills}  of desorption of
low energy Ps from Al, led to the estimate ${\sl s}(0)\approx 1$ by
means of a detailed balance argument. Because of the low mass of Ps,
quantum reflection effects are expected to be much more pronounced than
for incident atoms and molecules. According to Ref. \cite{martin2}, they
would normally be expected at incident energies below 2 eV. The
experiments explore energies down to $5\times 10^{-3}$ eV without signs
of quantum reflection.

Martin \etal\refto{martin2} offer as explanation their previous
conclusion  from numerical work that sufficiently strong inelastic
coupling eliminates quantum reflection. Our model shows quantum
reflection for neutral particles regardless of the strength of the
coupling. We also want to point out that the numerical results of Ref.
\cite{martin2} (Fig. 2) for a \underbar{charged} particle show that
${\sl s}(0)\propto \lambda^2$ for small coupling constant $\lambda$, in
line with our own results, and that, when plotted against $\lambda$
rather than $\log \lambda$, there is in our opinion no sign of an abrupt
transition. We therefore consider the apparent absence of quantum
reflection observed by Mills {\etal} as still unexplained.

We acknowledge helpful conversations with M.E. Flatt\'e, A. Kvinsinskii,
Th. Martin, and H. Metiu. This work was supported by NSF grant
DMR87-03434 and ONR grant N00014-89-J-1530.

\references

%EXPERIMENT
\refis{nayak} {V.U. Nayak, D.O. Edwards, and N. Masuhara,  Phys. Rev.
Lett. {\bf 50}, 990 (1983).}

\refis{schlichting}{H. Schlichting {\etal},  Phys. Rev. Lett. {\bf 60},
2515 (1988).}

\refis{mills} {A.P. Mills \etal, Phys. Rev. Lett. {\bf 66}, 735 (1991).}

\refis{doyle} {J.M. Doyle \etal, Phys. Rev. Lett. {\bf 67}, 603 (1991).}

%THEORY
\refis{lj&s} {J.E. Lennard--Jones and C. Strachan, Proc. R.
Soc. London, Ser. A {\bf 150}, 442 (1935).}

\refis{lj} {J.E. Lennard--Jones and A.F. Devonshire, Proc. Roy. Soc.
London, Ser. A {\bf 156}, 6 (1936).}

\refis{suhl} {T.R. Knowles and H. Suhl, Phys. Rev. Lett. {\bf 39}, 1417
(1977); {\bf 40}, 911E (1978).}

\refis{martin1} {Th. Martin, R. Bruinsma, and P.M. Platzman, Phys. Rev.
B {\bf 38}, 2257 (1988); {\bf 39}, 12411 (1989);{\bf 41}, 3172 (1990).}

\refis{martin2} {Th. Martin, R. Bruinsma, and P.M. Platzman, Phys. Rev.
B  {\bf 43}, 6466 (1991).}

\refis{ck} {D.P. Clougherty and W. Kohn, UC Santa Barbara preprint
UCSBTH-92-10 (submitted).}

\refis{brenig} {W. Brenig, Z. Phys. B {\bf 36}, 227 (1980).}

\refis{boheim} {J.~B\"oheim, W. Brenig, and J. Stutzki, Z. Phys. B {\bf
48}, 43 (1982).}

\refis{carraro} {C. Carraro and M.W. Cole, Phys. Rev. Lett. {\bf 68},
412 (1992).}

\refis{iche} {G. Iche and Ph. Nozi\'eres, J. Phys. (Paris) {\bf 37},
1313 (1976).}

\refis{stiles2} {M.D. Stiles, J.W. Wilkins, and M. Persson, Phys. Rev.
B {\bf 34}, 4490
 (1986)}.

\refis{goldberger} {M.L. Goldberger and K.M. Watson, {\bf Collision
Theory}  (R.E. Krieger, New York, 1975).}

\endreferences
\beginparmode
\noindent{Figure 1.}  (a) Schematic view of a particle with mass $m$
impinging upon a 1D solid consisting of atoms with mass $M$. Lattice
atoms are coupled to nearest neighbors and to fixed lattice sites. (b)
Particle interacts with  the end atom  via a finite-range surface
potential.

\end